\documentclass[a4paper,fleqn,usenatbib,useAMS]{mnras}
\usepackage{graphicx}	
\usepackage{amsmath}	
\usepackage{amssymb}	
\usepackage{multicol}        
\usepackage{bm}		
\usepackage{pdflscape}	
\usepackage[T1]{fontenc}
\usepackage{ae,aecompl}
\usepackage{color}
\title {{Clustering of Galaxies with Dynamical Dark Energy}}
 \author[B. Pourhassan et al.]{Behnam Pourhassan$^{1}$\thanks{E-mail: b.pourhassan@du.ac.ir} Sudhaker Upadhyay,$^{2}$\thanks{E-mail: sudhakerupadhyay@gmail.com}
  Mir  Hameeda,$^{3}$\thanks{E-mail: hme123eda@gmail.com}
 Mir Faizal$^{4}$\thanks{E-mail: mirfaizalmir@googlemail.com}\\
$^{1}$School of Physics, Damghan University, Damghan, Iran\\
$^{2}$Centre for Theoretical Studies, Indian Institute of Technology Kharagpur,  Kharagpur-721302,  India\\
$^{3}$Department of Physics, S.P. Collage,  Srinagar, Kashmir, 190001, India\\
$^{3}$ Visiting Associate, IUCCA,  Pune,  41100, India\\
$^{4}$Irving K. Barber School of Arts and Sciences, University of British Columbia - Okanagan,\\
3333 University Way, Kelowna,   British Columbia V1V 1V7, Canada\\
$^{4}$ Department of Physics and Astronomy, University of Lethbridge, Lethbridge, Alberta, T1K 3M4, Canada}

\begin{document}

\maketitle

\begin{abstract}
In this paper, we study thermodynamics of the cluster of galaxies under the effect of  dynamical dark energy. We evaluate the configurational integral for interacting system of galaxies in an expanding universe by including the effects produced by the varying $\Lambda$. The  gravitational partition function  is obtained using this configuration integral. We obtain thermodynamics quantities in canonical ensemble which depend on time and investigate the second law of thermodynamics. We also calculate the distribution function in grand canonical ensemble. The time evolution of the clustering parameter of galaxies is investigated for the time dependent (dynamical) dark energy. We conclude that the second law of thermodynamics is valid for the total system of cluster of galaxies and dynamical dark energy. We calculate correlation function and show that our model is very close to Peebles's power law, in agreement with the N-body simulation. It is observed that thermodynamics quantities  depend on the modified clustering parameter for this system of galaxies.
\end{abstract}

\begingroup
\let\clearpage\relax
\endgroup
\newpage

\section{Introduction}
Accelerated expansion of the Universe is established by the observations of the Type Ia Supernovae (SNeIa) (\cite{Riess}) and the Cosmic Microwave Background
(CMB) radiation anisotropies (\cite{Jarosik}). This accelerated expansion of the Universe can be explained by dark energy. In that case, there are several models to describe dark energy like $\Lambda$-CDM which tells that the cosmological constant $\Lambda$ plays the role of the dark energy (\cite{Bahcall}). However, this model has some famous ambiguities such as the fine-tuning (\cite{Copeland}), and the cosmic coincidence problems (\cite{Nobbenhuis}). Also, it is not a dynamical model of dark energy and cannot show evolution of dark side of the Universe. But, there are some dynamical model of dark energy like phantom (\cite{Caldwell}) quintessence (\cite{Wetterich}), K-essence (\cite{Armendariz-Picon}) and tachyonic models (\cite{Sen}), which are based on scalar fields. Also, there is an exotic fluid with a special equation of state called as Chaplygin gas (\cite{Kamenshchik}; \cite{Bento}) and its generalizations (\cite{Bilic}), modifications (\cite{Debnath}; \cite{Saadat2}), and extensions (\cite{Kahya}; \cite{Pourhassan}) which play   role of the dark energy and dark matter, and emerged initially in cosmology from string theory point of view (\cite{Bar1}; \cite{Bar2}). There are also some proposals of the holographic dark energy
(\cite{Li}; \cite{Elizalde}). $f(R)$ theories of gravity are alternative to dark energy to explain accelerating expansion of the Universe (\cite{Capozziello1}; \cite{Capozziello2}; \cite{Carroll}; \cite{Khur}). In order to have a dynamical model of dark energy, it is also possible to consider varying $\Lambda$ (\cite{Kahya}; \cite{Khur}; \cite{Jamil2}). In this paper, we will analyze the effect of time-dependent cosmological constant on the clustering of
galaxies. It is known that the inter-galactic distances are many orders of magnitudes greater than the length scale of a galaxy. Thus, it possible to represent a system of galaxies as a system of point particles, and analyze this system using standard methods of statistical mechanics (\cite{ahm02}). It is also possible to analyze the thermodynamics limit of such a system of interacting galaxies, and obtain appropriate thermodynamics
quantities for this system. In fact, the observed peculiar velocity distribution function has been studied for a sample of galaxies within $50 Mpc (H=100)$ of the local group (\cite{1a}). This was used to study the clustering parameters of this sample of galaxies.
A system of galaxies with halos has also been analyzed using these methods, and the peculiar velocity distribution function of such a system
of galaxies was analyzed using standard methods of statistical mechanics (\cite{2a}). The statistical mechanics has also been used to analyze spatial distribution function of galaxies at high redshift (\cite{4a}). The statistical mechanics has also been used in obtaining the probability to have certain shape of a galaxy cluster  (\cite{5a}). A system of two different kinds of galaxies have also been studied using
such statistical methods (\cite{6a}).  It may be noted that it is also possible to consider the effects generated by the finite size of galaxies using these statistical methods (\cite{7a}). Thus, the important point to be noted from these works is that it is
possible to use the standard techniques of statistical mechanics, to study the clustering of a system
of galaxies.
In this paper, we analyze the effect of the dynamical dark energy on the clustering of galaxies using the same standard methods of statistical mechanics. Already, Ref. \cite{main} discuss the effects of cosmological constant as dark energy on clustering of galaxies, but as explained above, cosmological constant is not dynamical model of dark energy and cannot show real evolution of dark side of the Universe. Hence, we consider a varying $\Lambda$ as dark energy and study dynamics of galaxies.\\

This paper is organized as follows. In the next section, we write the partition function and modified potential for the galaxies under effect of dynamical dark energy. Then, we extract thermodynamics quantities in section 3. The grand canonical ensemble will be discussing in section 4, and in section 5 we prove validity of the second law of the thermodynamics in presence of dark energy. In order to obtain agreement with observational data, we
calculate correlation function in section 6 and compare results with observations. Finally, in section 7, we present conclusion.
\section{Statistics}
\begin{figure}
\begin{center}$
\begin{array}{cccc}
\includegraphics[width=55 mm]{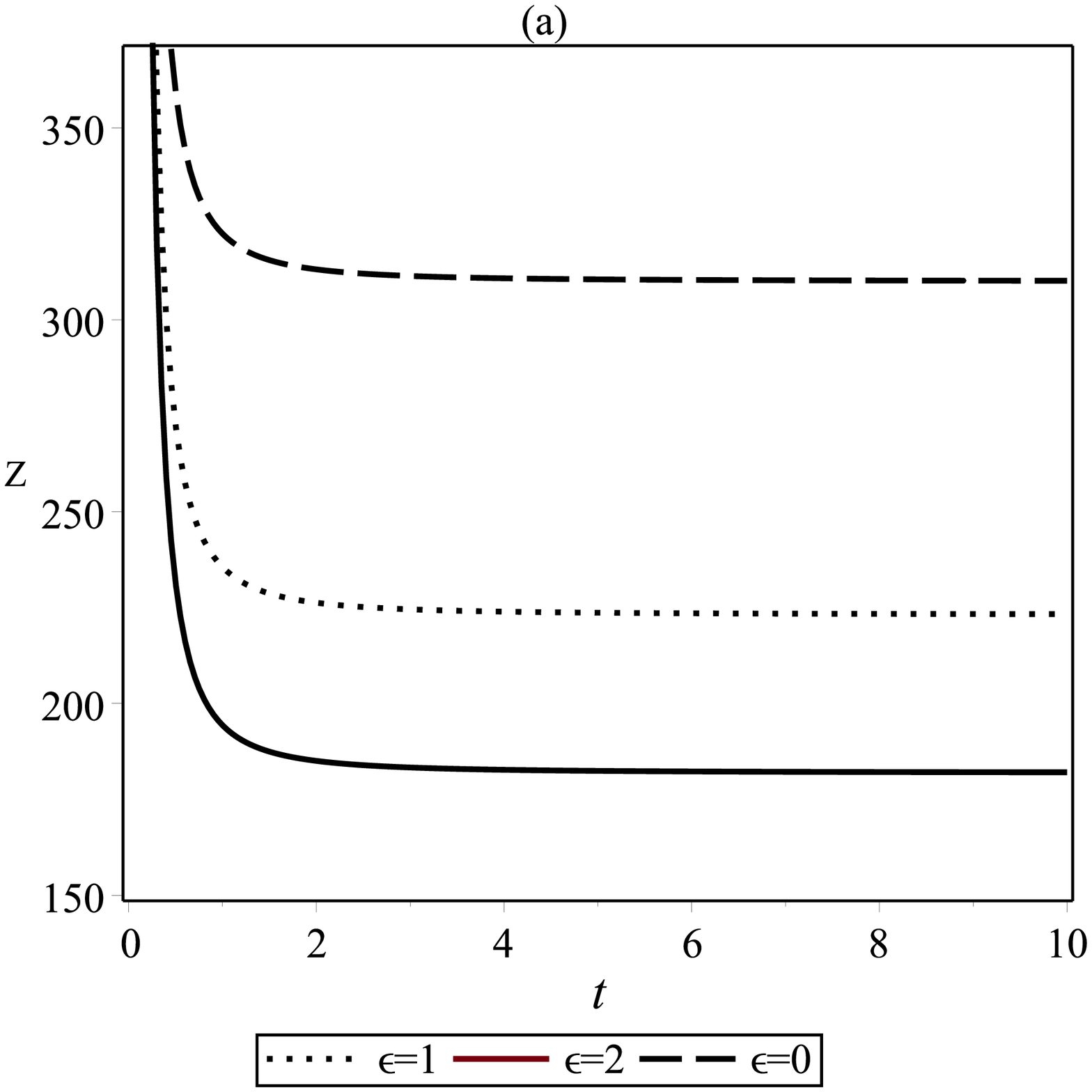}\\
\includegraphics[width=55 mm]{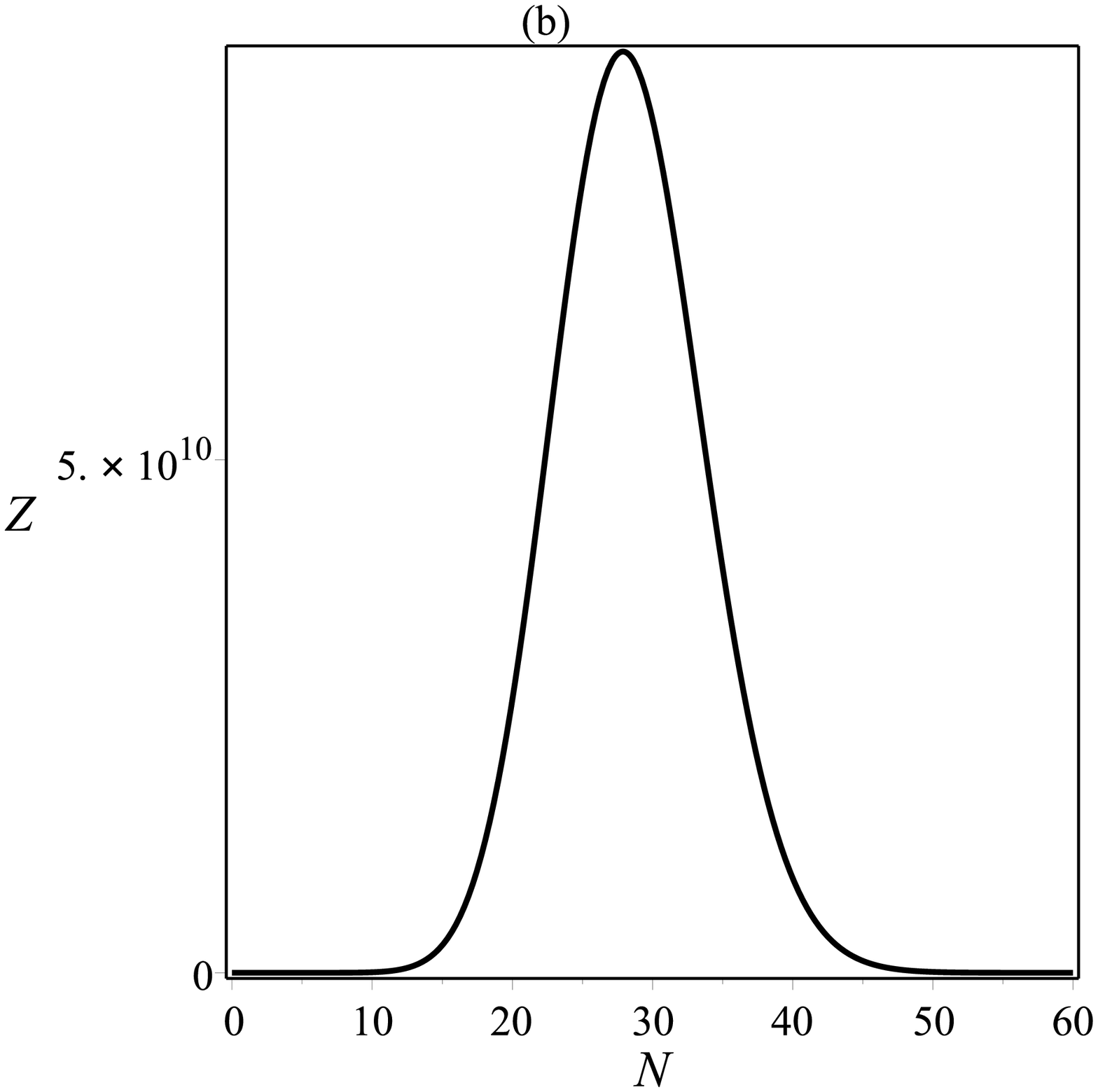}
\end{array}$
\end{center}
\caption{Typical behavior of the partition function in terms of $t$ (a) and $N$ (b) for $\tau=2$. We set unit value for all other parameters.}
\label{fig1}
\end{figure}

Partition function is an important statistical quantity which gives all thermodynamics information of the given system. Our interest here is to deal a system of galaxies which may be considered as collection of particles with pairwise interaction. From statistical point of view, we assume homogeneous distribution over the large region. Therefore, we consider the following partition function (\cite{ahm02}):
\begin{equation}\label{1}
Z(T,V)=\frac{1}{\lambda^{3N}N!}\int d^{3N}pd^{3N}r X,
\end{equation}
with the following definition:
\begin{equation}\label{1-1}
X\equiv \exp\biggl(-\frac{1}{T}\biggl[\sum_{i=1}^{N}\frac{p_{i}^2}{2m}+\Phi(r_{1}, r_{2}, r_{3}, \dots, r_{N})\biggr]\biggr),
\end{equation}
where $N$ is number of galaxies with mass $m$ and momenta $p_{i}$ interacting gravitationally with a potential $\Phi$, hence $N!$ takes the distinguishability of considered classical particles. Here, $T$ denotes average temperature, and $\lambda$ denotes the normalization factor. Also, we assume the unit value for the Boltzmann constant ($k=1$). The gravitational potential energy due to all pairs of particles composing the system, which is a function of the relative position vector $r_{ij}=|r_{i}-r_{j}|$, can be expressed as the sum of the potential of all pairs. It means that we can write $\Phi(r_{1}, r_{2}, \dots, r_{N})=\sum_{1\le i<j\le N}\phi(r_{ij})=\Phi(r_{ij})$. In that case, integration over the momentum space changes the partition function (\ref{1}) to the following expression:
\begin{equation}\label{zn}
Z_N(T,V)=\frac{1}{N!}\left(\frac{2\pi mT}{\lambda^2}\right)^{3N/2}Q_N(T,V),
\end{equation}
where
\begin{equation}\label{q1}
Q_{N}(T,V)=\int....\int \prod_{1\le i<j\le N} (f_{ij}+1)d^{3N}r.
\end{equation}
Here,  interaction function $f_{ij}$ is defined by
\begin{equation}\label{fun}
f_{ij}+1=e^{-\frac{\Phi(r_{ij})}{T}},
\end{equation}
where $f_{ij}=0$ denotes absence of interactions.
 It is found that the partition function diverges at $r_{ij}=0$ for the point particles (\cite{ahm02}).
This divergence can be removed if we consider extended nature of particles (particles with halo) (\cite{Saadat}), and it can be done via introducing a softening parameter in the Newtonian potential between particles as follow (\cite{ahm02}),
\begin{equation}\label{5}
\Phi(r_{ij})=-\frac{Gm^2}{(r_{ij}^2+\epsilon^2)^{1/2}},
\end{equation}
where $\epsilon$ denotes the softening parameter. However, dark energy may affect clustering of the galaxies by assuming cosmological constant $\Lambda$ as dark energy, which  modifies the potential as \cite{pe},
\begin{equation}\label{6}
\Phi(r_{ij})=-\frac{Gm^2}{(r_{ij}^2+\epsilon^2)^{1/2}}-\frac{\Lambda(t) r_{ij}^2}{6}.
\end{equation}
The case of constant $\Lambda$, in agreement with $\Lambda$-CDM model, has been studied recently by \cite{main}. However, we know that the dark energy is not constant and has dynamics, hence in order to have a real model, we consider time-dependent $\Lambda$. In another word, in order to consider effect of dynamical dark energy, a possible way is to consider the varying $\Lambda$ with time. This is justified in the equation (\ref{6}), where we presented a general form of potential energy with $\Lambda(t)$. For small variation in time, we can still assume the equilibrium thermodynamics.\\
For $\Phi(r_{ij})\ll T$ given in (\ref{6}), one can use Taylor expansion and the equation (\ref{fun}) in presence of dynamical dark energy can be rewritten as follow,
\begin{eqnarray}\label{7}
f_{ij}=\frac{Gm^{2}}{(r_{ij}^{2}+\epsilon^{2})^{1/2}T}+\frac{\Lambda(t) r_{ij}^2}{6T}.
\end{eqnarray}
Now, we can evaluate integral of the relation (\ref{q1}) over a spherical volume of radius $R_{1}$ and find that,
\begin{equation}\label{qn}
Q_{N}(T,V)=V^N\left[1+Y x\right]^{N-1},
\end{equation}
with the following definition:
\begin{equation}\label{qn-1}
Y\equiv\sqrt{1+\frac{\epsilon^2}{R_1^2}} +\frac{\epsilon^2}{R_1^2} \log \frac{{\epsilon/R_1}}{\left[ 1+\sqrt{1+\frac{\epsilon^2}{R_1^2}}\right]}+\frac{\Lambda(t) R_1^3}{15 Gm^2}.
\end{equation}
Here we used $R_{1}\sim \rho^{-1/3}\sim (\bar N/V)^{-1/3}$ and the following change of variable,
\begin{equation}\label{sc}
x\equiv\frac{3}{2}\left(\frac{ Gm^{2}}{T}\right)^3\rho,
\end{equation}
and we utilized the following scale transformations (\cite{ahm02}; \cite{sas01}),
\begin{eqnarray}\label{10}
\rho&\rightarrow& \lambda^{-3}\rho,\nonumber\\
T&\rightarrow&\lambda^{-1}T,\nonumber\\
R_1&\rightarrow& \lambda R_1,\nonumber\\
\frac{Gm^2}{R_1T}&\rightarrow& \left(\frac{Gm^2}{R_1T}\right)^{3}.
\end{eqnarray}
Hence, the gravitational partition function is given by,
\begin{eqnarray}\label{11}
&&Z_N(T,V)=\frac{V^{N}}{N!}\left(\frac{2\pi mT}{\lambda^2}\right)^{3N/2}\nonumber\\
&&\times\left[1+\frac{3}{2}Y \left(\frac{ Gm^{2}}{TR_{1}}\right)^3\right]^{N-1}.
\end{eqnarray}
In the simplest way, we can consider the following time-dependence,
\begin{equation}\label{12}
\Lambda(t)\propto\frac{1}{t^{\tau}},
\end{equation}
where $\tau$ is a positive constant and can be fixed using observational data. However, without loose of generality, we can set $\tau=2$ to see typical behavior of thermodynamics and statistics quantities (finally we show that $\tau=2$ is the best fitted value).
In the Fig. \ref{fig1}, we draw partition function and see that it is decreasing function of time as expected (see Fig. \ref{fig1} (a)). Variation of partition function is high at initial time and then leads to a constant corresponding to the current value. We also show that the softening parameter decreases the value of the partition function. In the Fig. \ref{fig1} (b), we draw partition function in terms of $N$ and see that it is an increasing function for small $N$, while it is a decreasing function for large $N$. There is a critical $N$ which corresponds to the maximum of partition function which means maximum of probability.\\
In the next section, we use the partition function (\ref{11}) to extract some thermodynamics quantities in canonical ensemble and study the effect of dynamical dark energy on them.

\section{Thermodynamics}
\begin{figure}
\begin{center}$
\begin{array}{cccc}
\includegraphics[width=55 mm]{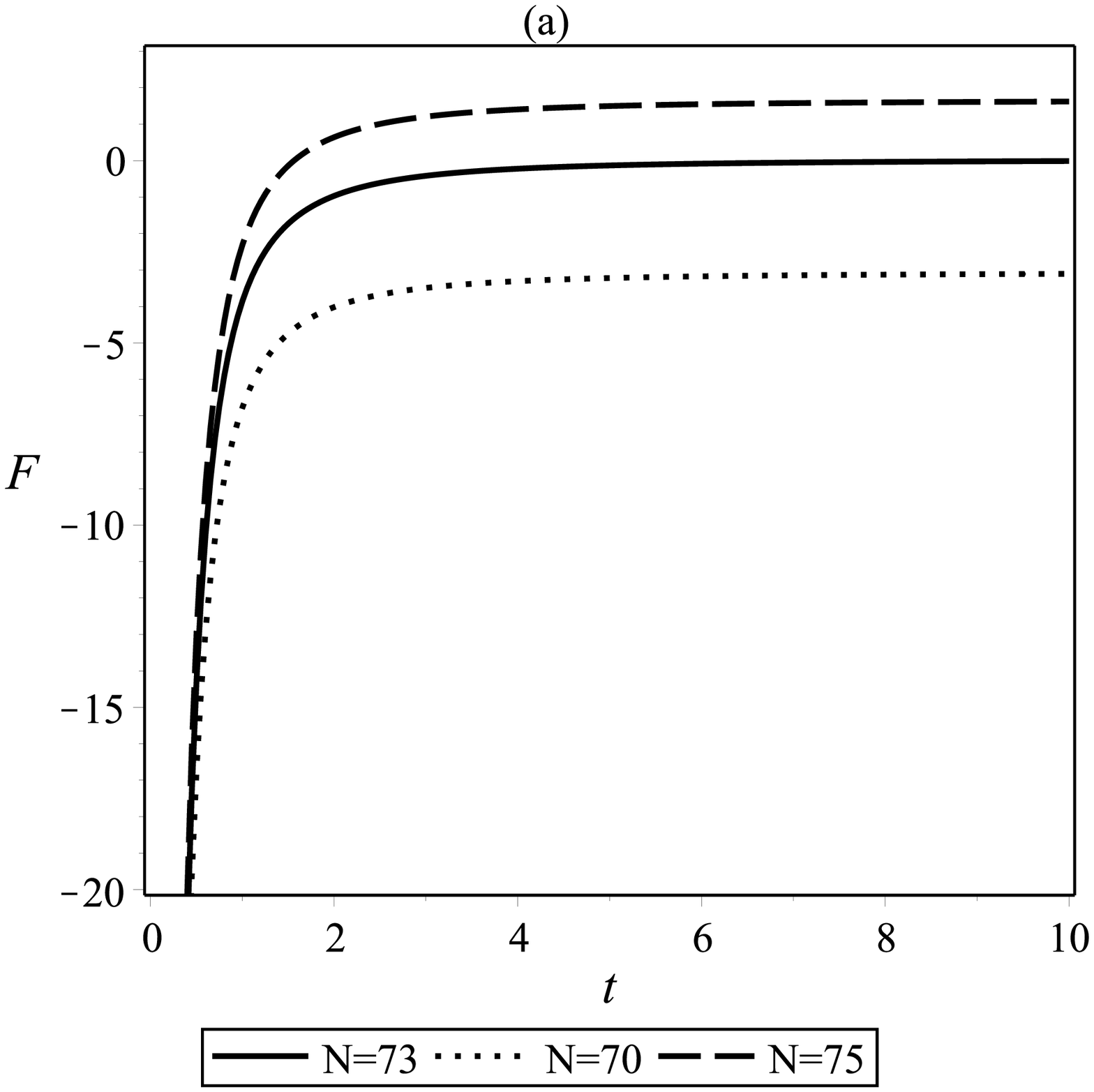}\\
\includegraphics[width=55 mm]{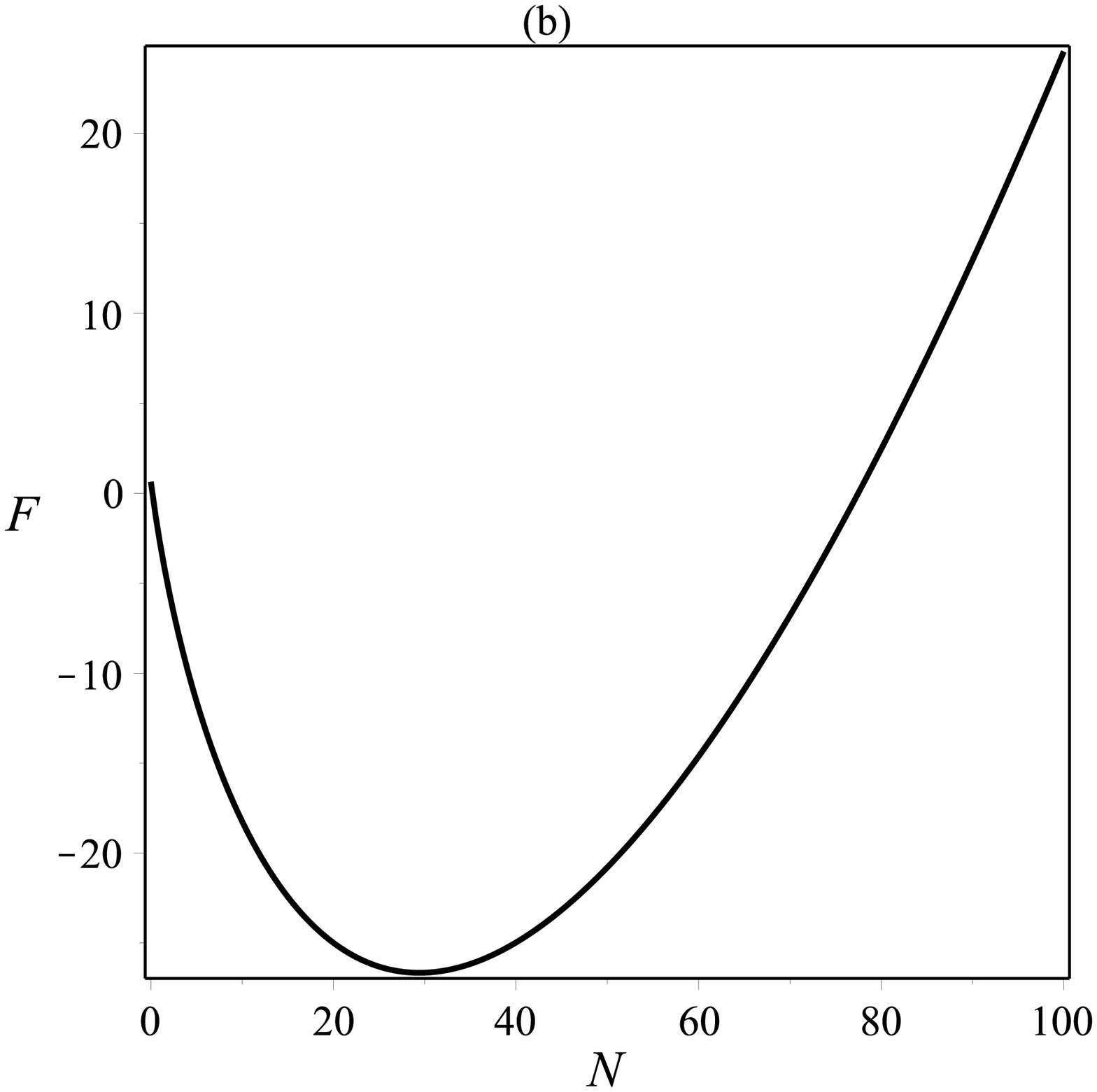}
\end{array}$
\end{center}
\caption{Typical behavior of the Helmholtz free energy in terms of $t$ (a) and $N$ (b) for $\tau=2$. We set unit value for all parameters.}
\label{fig2}
\end{figure}

\begin{figure}
\begin{center}$
\begin{array}{cccc}
\includegraphics[width=55 mm]{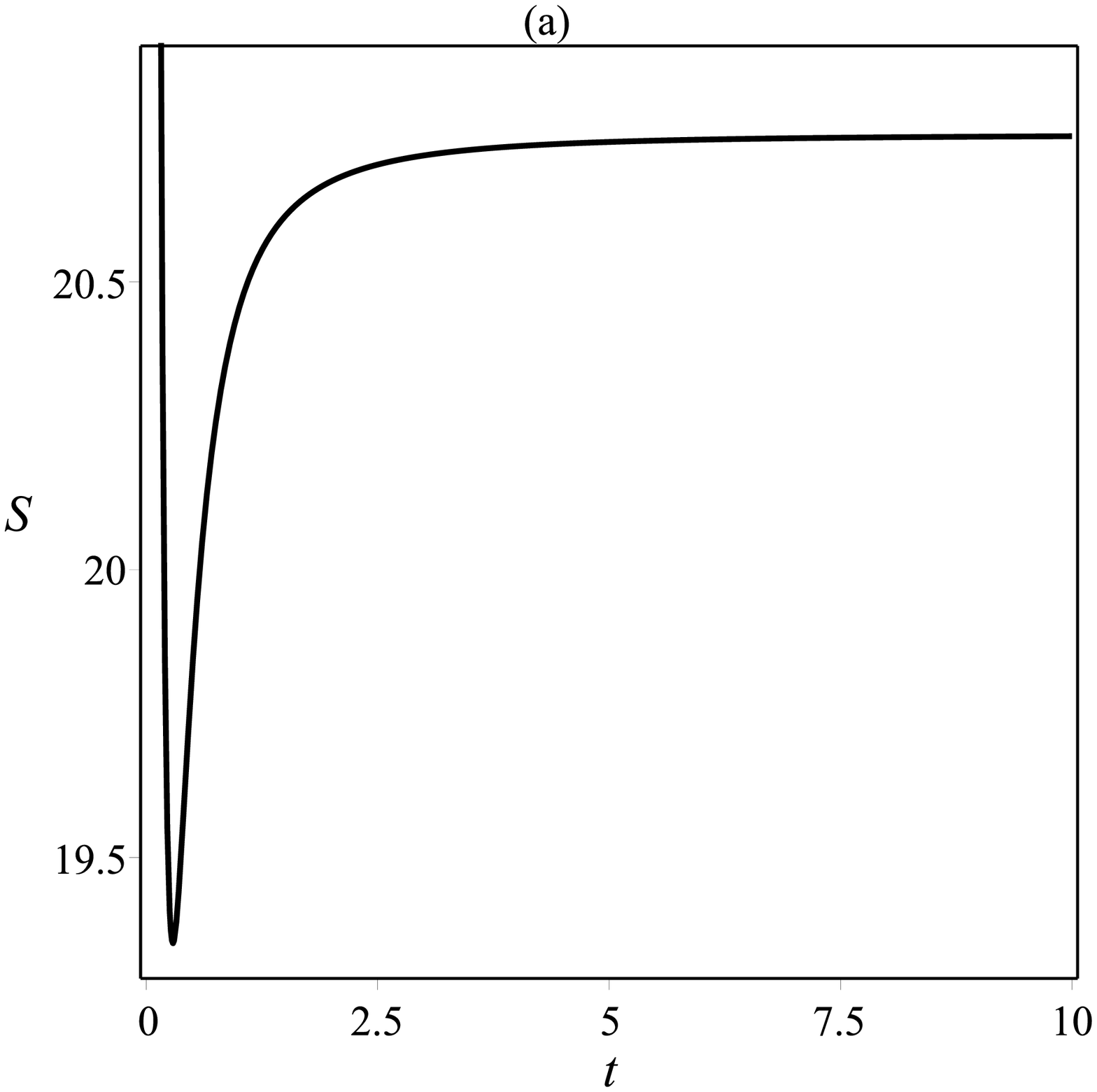}\\
\includegraphics[width=55 mm]{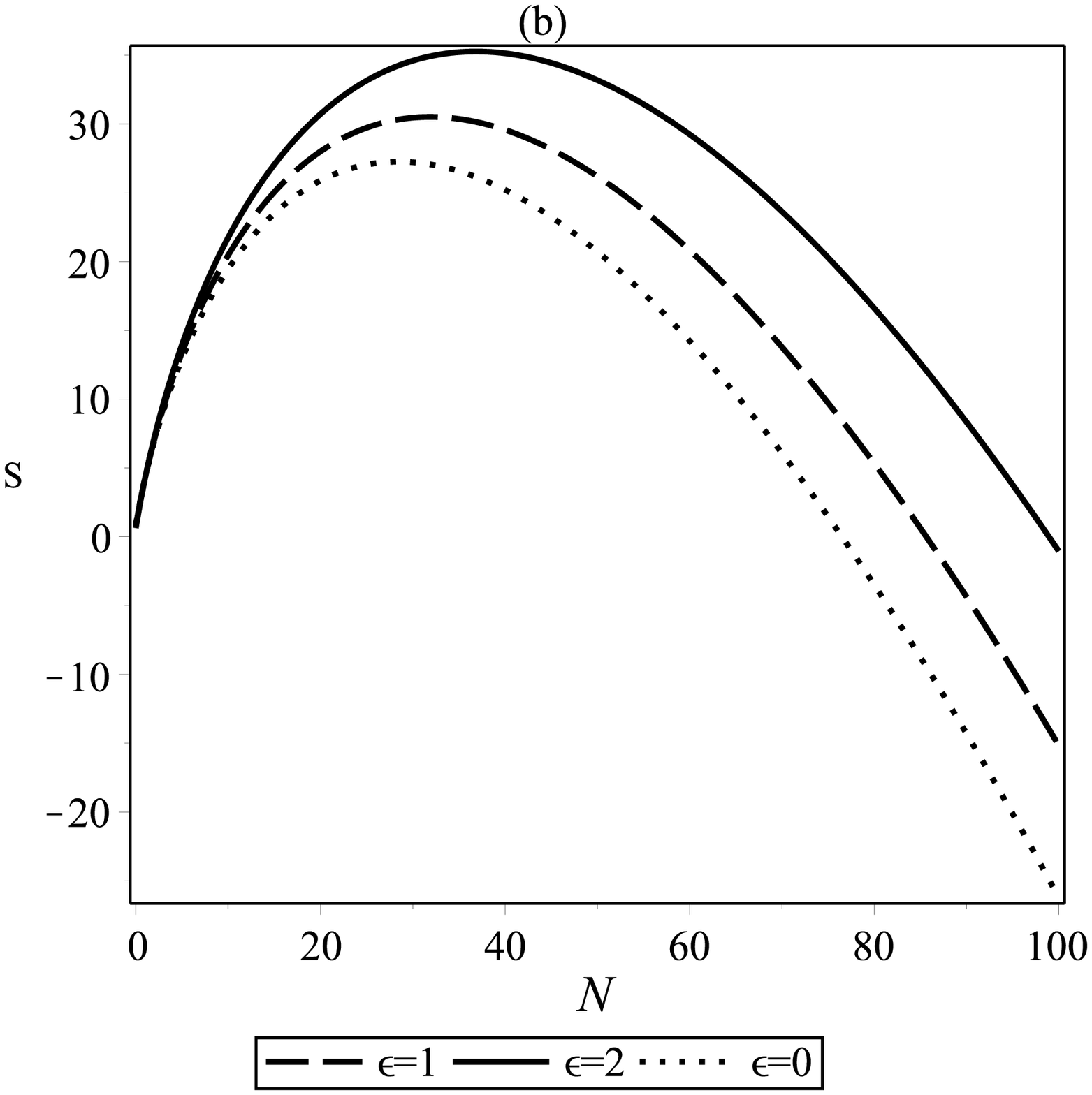}
\end{array}$
\end{center}
\caption{Typical behavior of the entropy in terms of $t$ (a) and $N$ (b) for $\tau=2$. We set unit value for all parameters.}
\label{fig3}
\end{figure}

\begin{figure}
\begin{center}$
\begin{array}{cccc}
\includegraphics[width=55 mm]{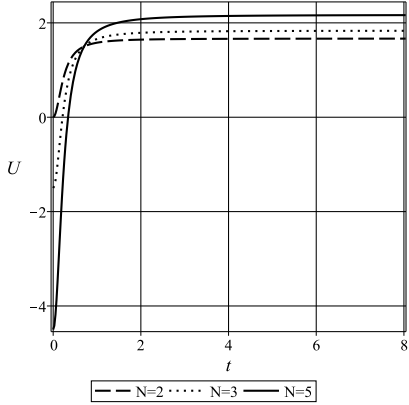}
\end{array}$
\end{center}
\caption{Typical behavior of the internal energy in terms of $t$ for $\tau=2$. We set unit value for all parameters.}
\label{fig4}
\end{figure}

\begin{figure}
\begin{center}$
\begin{array}{cccc}
\includegraphics[width=55 mm]{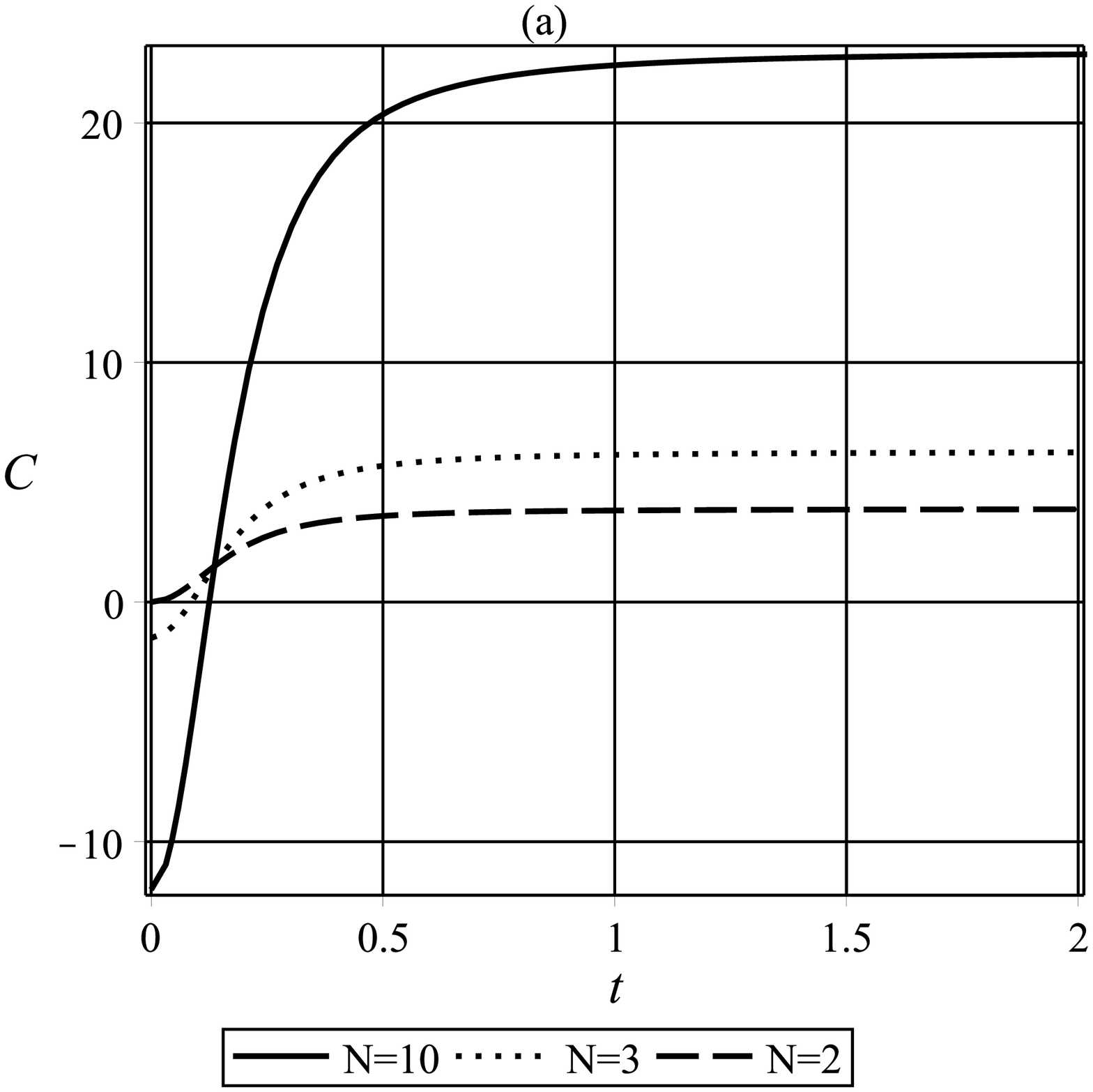}\\
\includegraphics[width=55 mm]{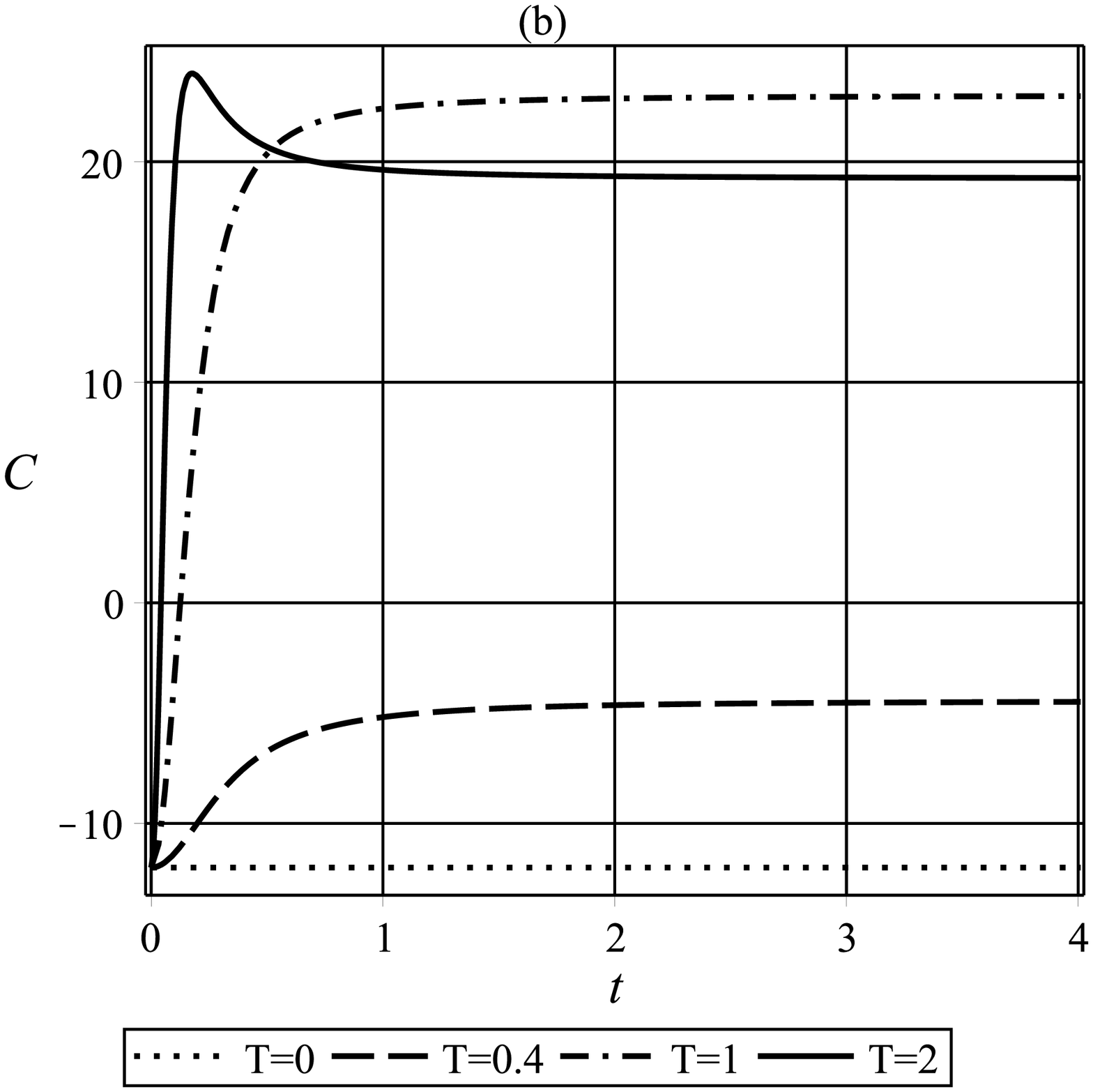}
\end{array}$
\end{center}
\caption{Typical behavior of the specific heat in terms of $t$ for $\tau=2$, (a) with variation of $N$, and (b) with variation of temperature for $N=10$. We set unit value for all other parameters.}
\label{fig5}
\end{figure}

Now, our goal is to calculate thermodynamic quantities of strongly interacting system of galaxies interacting through a modified
Newtonian potential which describes the galactic clustering   affected by the dynamical dark energy.
Helmholtz free energy is one of the important thermodynamics quantity which provides the normalization factor for the probability distribution. In the canonical ensemble, Helmholtz free energy is given by,
\begin{equation}\label{13}
F=-T\ln Z_{N}(T,V).
\end{equation}
Hence, we can obtain,
\begin{equation}\label{f}
F=-T\ln\left(\frac{V^N}{N!}\left(\frac{2\pi mT}{\lambda^2}\right)^{3N/2}\left[1+\frac{3}{2}Y\left(\frac{ Gm^{2}}{TR_{1}}\right)^3\right]^{N-1}\right).
\end{equation}
In the Fig. \ref{fig2}, we draw Helmholtz free energy for $\tau=2$, and see that it is increasing function of time as illustrated by Fig. \ref{fig2} (a). The value of the Helmholtz free energy will be positive for sufficient large $N$ as illustrated by the Fig. \ref{fig2} (b). We can obtain similar behavior for another value of $\tau$ as well.
Then, we can use the following relation to compute entropy:
\begin{equation}\label{15}
S= -\biggl(\frac{\partial F}{\partial T}\biggr)_{N,V}.
\end{equation}
In that case one can find,
\begin{eqnarray}\label{s}
S&=&N\ln\left(\frac{V}{N}T^{3/2}\right)+N\ln\left(\frac{2\pi m}{\lambda^2}\right)\nonumber\\
&+&(N-1)\ln\left[1+\mathcal{A}\right]-3N\mathcal{B}+\frac{5}{2}N+\frac{3}{2},
\end{eqnarray}
where we defined,
\begin{equation}\label{17}
\mathcal{A}=\frac{3}{2}Y\left(\frac{ Gm^{2}}{TR_{1}}\right)^{3},
\end{equation}
and clustering parameter of galaxies defined by \cite{ahm02},
\begin{equation}\label{b}
\mathcal{B}=\frac{\mathcal{A}}{\left[1+\mathcal{A}\right]}.
\end{equation}
It is indeed important parameter which plays a key role to find various thermodynamic quantities. It is obvious that $0\leq\mathcal{B}\leq1$ and it is a decreasing function of $\varepsilon$, while   increasing function of $\Lambda(t)$. At the initial time ($t\ll1$),
 it is clear that $\mathcal{A}\gg1$ and one can find $\mathcal{B}\approx1$, and entropy diverges at the initial time as illustrated by the Fig. \ref{fig3} (a). The case of $\mathcal{B}=1$ happens only if we consider varying dark energy which is consistent with the nature of our universe. We can see that at the initial time, entropy is a decreasing function of time, which indicates that the second law of thermodynamics
 is violated. However, in order to study the second law of thermodynamics, we should also consider the dynamical dark energy density (it is subject of the next section). Also, in the Fig. \ref{fig3} (b), we can see the variation of the entropy with $N$ and can find critical $N$ as before to have maximum of the entropy corresponding to the equilibrium. From the Fig. \ref{fig3} (b), it is clear that the entropy enhanced due to the effect of $\varepsilon$. It means that the entropy corresponding to the extended masses is larger than the entropy corresponding to the point masses.\\
The clustering parameter for  $\epsilon =0$, which  corresponds to the point masses,
is given by
\begin{equation}\label{19}
\mathcal{B}_0  =  b \left( \frac{{15 Gm^{2}t^{\tau}} + {R_1^3} }{ {15 Gm^{2}t^{\tau}} +b {R_1^3} } \right),
\end{equation}
where $b= \frac{x}{1+ x}$ denotes the original clustering parameter of Newtonian potential for point masses (\cite{ahm02}), and $\mathcal{B}_0$ is the corrected clustering parameter for the point masses due to the dynamical dark energy. After quite long time (i.e., $15 Gm^{2}t^{\tau}\gg b {R_1^3}$) $\mathcal{B}_0  =  b$ as expected.\\
By using the relations (\ref{f}) and (\ref{s}), one can obtain internal energy of a system of galaxies as,
\begin{equation}\label{20}
U=F+TS=\frac{3}{2}NT\big(1-2\mathcal{B}\big).
\end{equation}
In the Fig. \ref{fig4}, we can see behavior of the internal energy with time. It has been shown that for the $N<2$ the internal energy is completely positive, while for the $N>2$ the value of the internal energy is negative at the initial time. We find that the internal energy is an increasing function of softening parameter. It means that internal energy of the extended masses is larger than the internal energy of the point masses.\\
The specific heat at constant volume is an important thermodynamics parameter, which can be calculated via the following relation:
\begin{equation}\label{21}
C=T\left(\frac{\partial S}{\partial T}\right)_{V}.
\end{equation}
In the Fig. \ref{fig5}, we can see the time evolution of the specific heat. Similar to the internal energy, we can see that specific heat is completely positive for the small $N$ ($N<2$), however there is some negative regions for the larger $N$ at the initial time where dark energy has large value. It means that the system is thermodynamically unstable for the large dark energy. Fig. \ref{fig5} (b) shows that the specific heat also depends on the temperature. For the low temperature case, there are some negative regions, while for the high temperature we have completely positive specific heat. There is a critical temperature where the specific heat has a maximum.\\

In the next section, we consider grand canonical ensemble to extract some more general relations.
\section{Grand canonical ensemble}
\begin{figure}
\begin{center}$
\begin{array}{cccc}
\includegraphics[width=55 mm]{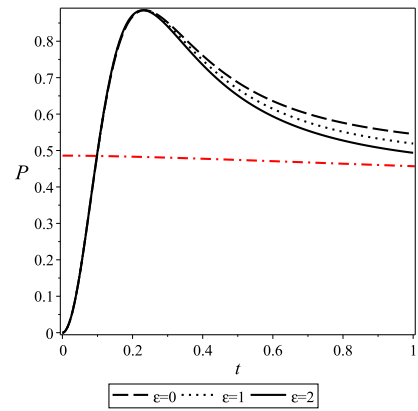}
\end{array}$
\end{center}
\caption{Typical behavior of the distribution function in terms of $t$ for $\tau=2$, $\bar{N}=10$ ($R_{1}=2$ and $V=80$) and $N=3$. We set unit value for all other parameters. Dash dotted line denotes the case of without dark energy.}
\label{fig6}
\end{figure}

Grand canonical partition function in terms of canonical partition function is given by,
\begin{eqnarray}\label{g}
Z_{G}(T,V,z)=\sum_{N=0}^{\infty}z^{N}Z_{N}(V,T),
\end{eqnarray}
where
\begin{eqnarray}\label{23}
z=e^{\frac{\mu}{T}}
\end{eqnarray}
is called  activity, where
\begin{eqnarray}\label{24}
\mu = \biggl(\frac{\partial F}{\partial N}\biggr)_{V,T},
\end{eqnarray}
is chemical potential. Hence, the grand partition function for the gravitationally interacting system of galaxies is obtained as,
\begin{eqnarray}\label{z}
Z_{G}(T,V,z)=e^{N(1-\mathcal{B})}.
\end{eqnarray}
By using the relation (\ref{z}), one can obtain the probability of finding $N$ particles in volume $V$ as follow,
\begin{eqnarray}\label{26}
P=\frac{e^{\frac{N\mu}{T}}Z_{N}(V,T)}{Z_{G}(T,V,z)}.
\end{eqnarray}
In the Fig. \ref{fig6}, one can see the time evolution of the probability with variation of the softening parameter. Here, we find that the difference of point and extended masses is
negligibly small at initial time. By decreasing dark energy to a constant value at the late time, we can see the difference of distribution constants corresponding to the point and extended masses. Also, the galaxy distribution function leads to a constant value at the late time. In absence of dark energy (see dash dot line of the Fig. \ref{fig6}), we can see that galaxy distribution function is approximately constant corresponding to the late time dark energy.
\section{Dark energy thermodynamics}
In this paper as well as previous work of \cite{main}, the effect of dark energy on the ordinary matter investigated. It means that we have an interacting dark energy model. In that case the evolution rate of the dark energy entropy for the dark energy density $\rho_{de}$
is given by \cite{setare},
\begin{eqnarray}\label{27}
{\dot{S}}_{de}=\frac{4\pi}{T}(1+\omega^{eff}_{de})\rho_{de} r_{A}^{2}(\dot{r}_{A}-Hr_{A}),
\end{eqnarray}
where $r_{A}$ is the apparent horizon which is associated with the gravitational entropy for a dynamical space-time and given by the following expression in terms of Hubble expansion parameter for the flat space-time (\cite{ch}):
\begin{eqnarray}\label{28}
r_{A}=H^{-1}.
\end{eqnarray}
The temperature, $T$, is given by,
\begin{eqnarray}\label{29}
T=\frac{1}{2\pi r_{A}}.
\end{eqnarray}
Also, the effective equation of state $\omega^{eff}_{de}$ satisfies the following conservation equation:
\begin{eqnarray}\label{30}
\dot{\rho}_{de}+3H(1+\omega^{eff}_{de})\rho_{de}=0,
\end{eqnarray}
and the Hubble expansion parameter can be written in terms of the scale factor $a$ as follow,
\begin{eqnarray}\label{31}
H=\frac{\dot{a}}{a}.
\end{eqnarray}
At the simplest model, we can consider the following scale factor:
\begin{eqnarray}\label{32}
a\propto t^{n},
\end{eqnarray}
which yields to the following Hubble expansion parameter:
\begin{eqnarray}\label{33}
H=\frac{n}{t}.
\end{eqnarray}
It is also possible to assume the following time-dependent equation of state (\cite{Gor}):
\begin{eqnarray}\label{34}
\omega^{eff}_{de}=\omega_{1}t+\omega_{0},
\end{eqnarray}
which appears due to the modification of gravity (\cite{noj}), where $\omega_{0}$ and $\omega_{1}$ are constants. Using this value of $\omega^{eff}_{de}$ in the equation (\ref{30}), we have a differential equation for the dark energy density with the following solution:
\begin{eqnarray}\label{35}
\rho_{de}=\rho_{0}\frac{e^{-c_{1}t}}{t^{-c_{2}}},
\end{eqnarray}
where constants $\omega_{0}$ and $\omega_{1}$ are encoded in the new constants $c_{1}=3n\omega_{1}$ and $c_{2}=-3n(1+\omega_{0})$, and $\rho_{0}$ corresponds to the constant dark energy density at $\omega_{0}=-1$ and $\omega_{1}=0$. In this scenario, the equation (\ref{27}) reduces to the following differential equation:
\begin{eqnarray}\label{36}
{\dot{S}}_{de}=(A_{1}t^{m+1}+A_{2}t^{m})e^{-c_{1}t},
\end{eqnarray}
where
\begin{eqnarray}\label{36-1}
A_{1}&=&\frac{4\pi(1-n)\omega_{1}\rho_{0}}{Tn^{3}},\nonumber\\
A_{2}&=&\frac{4\pi(1-n)(1+\omega_{0})\rho_{0}}{Tn^{3}},\nonumber\\
m&=&-3n(1+\omega_{0})+2.
\end{eqnarray}

\begin{figure}
\begin{center}$
\begin{array}{cccc}
\includegraphics[width=50 mm]{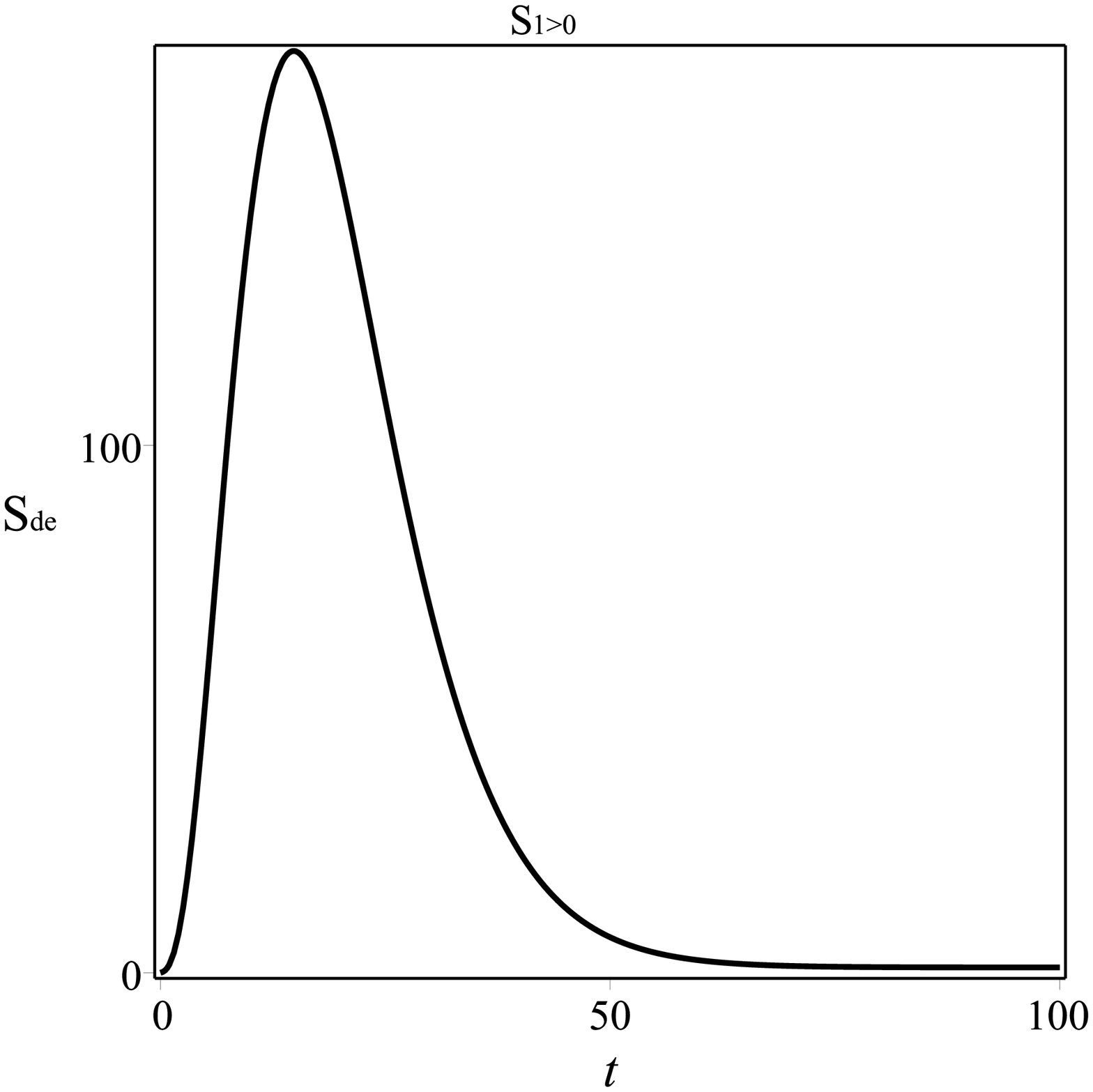}\\
\includegraphics[width=50 mm]{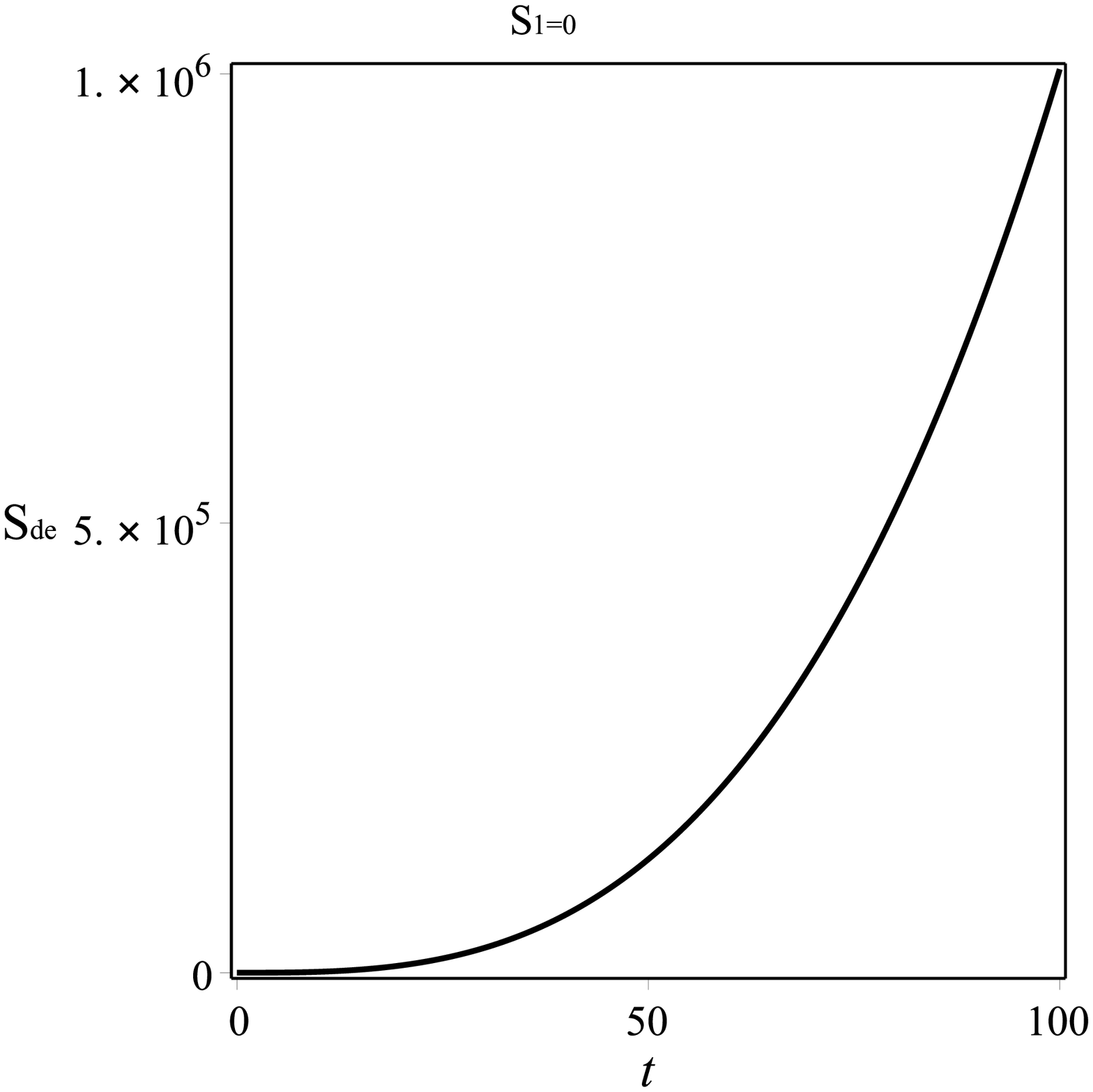}\\
\includegraphics[width=50 mm]{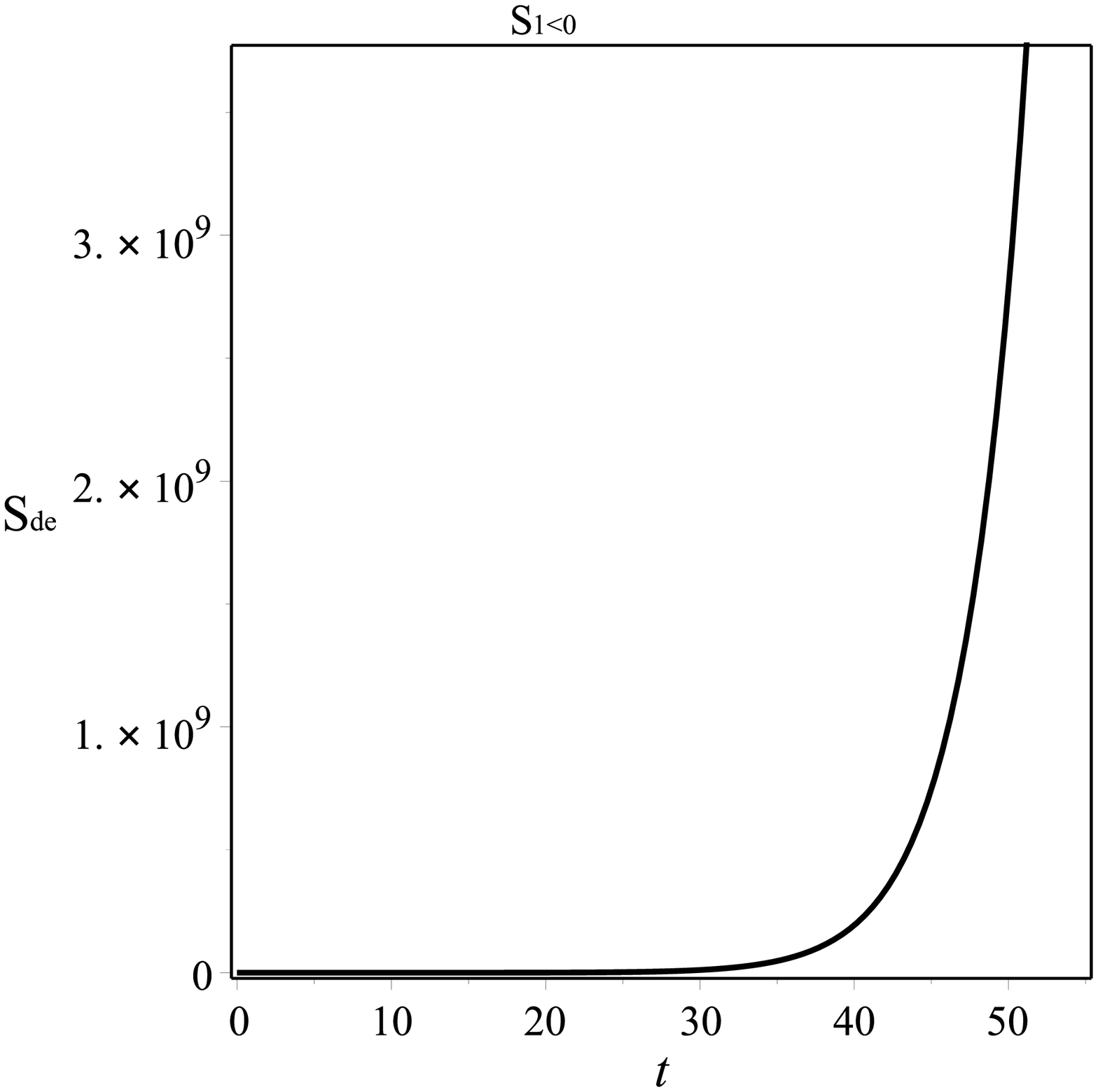}
\end{array}$
\end{center}
\caption{Typical behavior of ${S}_{de}$ in terms of $t$ for $m=2$ and with $S_{0}=S_{2}=S_{3}=1$.}
\label{fig7}
\end{figure}

For example, for the $m=2$ (eg. $n=1$, $\omega_{0}=-1$), we have $c_{2}=0$ and equation (\ref{36}) yields to the following equation:
\begin{eqnarray}\label{37}
{\dot{S}}_{de}=(A_{1}t^{3}+A_{2}t^{2})e^{-c_{1}t}.
\end{eqnarray}
Equation (\ref{37}) has the following general solution with the fact that value of the dark energy entropy is zero at the $t=0$:
\begin{eqnarray}\label{38}
{S}_{de}=S_{0}+(S_{3}t^{3}+S_{2}t^{2}+S_{1}t-S_{0})e^{-c_{1}t},
\end{eqnarray}
where $S_{i}$ are some constants involve $c_{1}$, $A_{1}$ and $A_{2}$ as follows,
\begin{eqnarray}\label{38-1}
S_{3}&=&-\frac{A_{1}}{c_{1}},\nonumber\\
S_{2}&=&-\frac{A_{2}}{c_{1}}-\frac{3A_{1}}{c_{1}^{2}},\nonumber\\
S_{1}&=&-\frac{2A_{2}}{c_{1}^{2}}-\frac{6A_{1}}{c_{1}^{3}},\nonumber\\
S_{0}&=&-\frac{2A_{2}}{c_{1}^{3}}-\frac{6A_{1}}{c_{1}^{4}}.
\end{eqnarray}
We can draw ${S}_{de}$ and find that the value of $S_{0}$, $S_{3}$ and $S_{2}$ must be positive to have positive entropy, while we can choose any value for the $S_{1}$ (see plots of the Fig. \ref{fig7}). For the positive $S_{1}$, we see a maximum entropy which means that after some time the value of the entropy decreases with time which violates the second law of thermodynamics. But for negative and zero $S_{1}$, we have completely increasing entropy with time.

\begin{figure}
\begin{center}$
\begin{array}{cccc}
\includegraphics[width=55 mm]{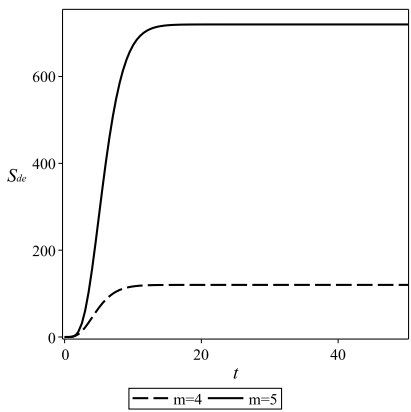}
\end{array}$
\end{center}
\caption{Typical behavior of ${S}_{de}$ in terms of $t$ for $m=4$ and $m=5$ with $A_{0}=c_{1}=1$.}
\label{fig8}
\end{figure}

In the general case with arbitrary $m$, the equation (\ref{36}) has the following solution:
\begin{eqnarray}\label{39}
{S}_{de}=A_{0} t^{\frac{m}{2}}e^{-c_{1}t}WM(\frac{m}{2},\frac{m+1}{2},c_{1}t),
\end{eqnarray}
where $WM(\mu,\nu,x)$ is the Whittaker function which can be defined in terms of the hypergeometric function as follow,
\begin{eqnarray}\label{40}
WM(\mu,\nu,x)=e^{-\frac{x}{2}}x^{\nu+\frac{1}{2}}F([\nu-\mu+\frac{1}{2}], [1+2\nu], x).
\end{eqnarray}
In the Fig. \ref{fig8}, we can see the typical behavior of  ${S}_{de}$ for $m=4$ and $m=5$, and find that the entropy is totally increasing function of time.

Now, we are in the position to study the total entropy due to the dark energy and cluster of galaxies, which is given by,
\begin{eqnarray}\label{41}
S_{tot}={S}_{de}+S,
\end{eqnarray}
where entropy of galaxies already given in the equation (\ref{s}), and ${S}_{de}$ is given in (\ref{38}) or (\ref{39}). ${S}_{de}$ denotes the dark energy entropy and $S$ denotes the entropy of galaxies under effect of dark energy which   indeed includes the interaction of matter with the dark energy. It is easy to check that the entropy (\ref{41}) is completely increasing function of time and the second law of the thermodynamics is valid.
\section{Dynamical evolution of galaxy-galaxy correlation function}
\begin{figure}
\begin{center}$
\begin{array}{cccc}
\includegraphics[width=55 mm]{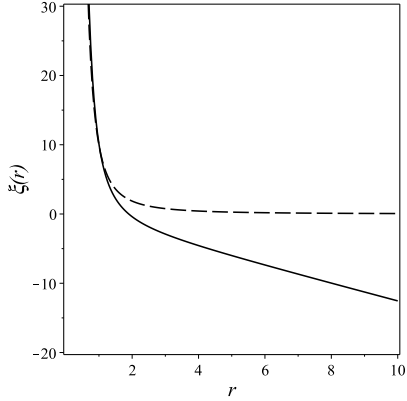}
\end{array}$
\end{center}
\caption{Typical behavior of galaxy correlation function $\xi(r)$ in terms of $r$ for $\epsilon=1$. Solid line represents the case with the dark energy while dashed line represents the case without the dark energy.}
\label{fig9}
\end{figure}

In order to show the agreement of our results with the observational data, we calculate the
correlation function.
The two-point correlation function $\xi(x)$ in clustering of galaxies obeys power law $\xi(x)=r^{-1.8}$ (\cite{pee80}) and the validity of this power law has been confirmed from
$N$-body simulation (\cite{sut90}).
In this section, we develop a power law for the correlation function using modified Newtonian potential in presence of dynamical dark energy.\\
The basic equation of energy $U$ for a spherical system of volume  $V$  consisting of $N$ particles is given by (\cite{sas00}; \cite{ham16}),
\begin{equation}\label{42}
U=\frac{3}{2}NT-\frac{N\rho}{2}\int_{V}\phi(r)\xi(r)4\pi r^2dr
\end{equation}
Using the interaction potential $\Phi$ between two galaxies in presence of dynamical dark energy, given in the equation (\ref{5}), and the  clustering parameter,
\begin{equation}\label{44}
\mathcal{B}=\frac{Gm^2\rho}{6T}\int\left[\frac{\xi(r)}{\sqrt{\epsilon^2+r^2}}+\frac{\Lambda(t) r^3}{6}\right]dV,
\end{equation}
we can obtain correlation function $\xi(r)$.
 The volume derivative of $\mathcal{B}$ is given as
\begin{eqnarray}\label{45}
\frac{\partial \mathcal{B}}{\partial V}&=&\frac{Gm^2\rho}{6T}\frac{\partial}{\partial V}\left[\int\left(\frac{\xi(r)}{\sqrt{\epsilon^2+r^2}}+\frac{\Lambda(t) r^3}{6} \right)
dV\right]\nonumber\\
&+&\frac{Gm^2}{6T}\frac{d\rho}{dV}\int\left(\frac{\xi(r)}{\sqrt{\epsilon^2+r^2}}+\frac{\Lambda(t) r^3}{6}\right)dV.
\end{eqnarray}
Using relation
\begin{equation}\label{46}
\frac{\partial \mathcal{B}}{\partial V}=\frac{\partial \rho}{\partial V}\frac{\partial \mathcal{B}}{\partial \rho},
\end{equation}
in the equation (\ref{45}), we obtain
\begin{equation}\label{47}
\frac{\partial \mathcal{B}}{\partial \rho}=\frac{\mathcal{B}(1-\mathcal{B})}{\rho}.
\end{equation}
From $\rho=\frac{N}{V}$, we have
\begin{equation}\label{48}
\frac{\partial \rho}{\partial V}=-\frac{\rho}{V}.
\end{equation}
Neglecting higher powers of $\epsilon/r$ and using above equations,
 we get the functional form of correlation function as,
\begin{equation}\label{49}
\xi(r)=\frac{9{\mathcal{B}}^{2}T}{2\pi Gm^2\rho}\frac{1}{r^2}\bigl(1+\frac{\epsilon^2}{2r^2}-\frac{\Lambda(t) r^3}{6}\bigr),
\end{equation}
and the time evolution of correlation function is expressed as,
\begin{equation}\label{50}
\xi(r)=\frac{9{\mathcal{B}}^{2}T}{2\pi Gm^2\rho}\frac{1}{r^2}\bigl(1+\frac{\epsilon^2}{2r^2}-\frac{r^3}{6t^{\tau}}\bigr).
\end{equation}
The equations (\ref{28}), (\ref{29}) and (\ref{33}) together suggest  that $t\propto r$,
and hence the correlation function (\ref{50}) is very close to Peebles's power for
$\tau\approx2$ which shows the impact of time evolution of dark energy. In the Fig. \ref{fig9} we can see effect of dark energy on the two-point
galaxy correlation function (\ref{49}) and find  that the effect of dark energy is important for larger value of $r$.

\section{Conclusion}
In this paper, we have analyzed the effect of dynamical dark energy on the cluster of galaxies by using the standard techniques of statistical mechanics, and found that two-point correlation function in presence of time-dependent dark energy is very close to Peebles's power law. It indicates that our analysis is also close to the actual physical system of interacting galaxies. We have considered varying $\Lambda$ as a function of time to make a dynamical model of dark energy and calculated time-dependent partition function in both canonical and grand canonical ensembles. This was done by modifying the partition function of a system of galaxies by a varying $\Lambda$ term. Indeed, we analyzed a model of
time-dependent dark energy for this system. Then, we obtained the effect of time-dependent dark energy on the thermodynamics quantities like Helmholtz free energy. We have shown that the free energy is an increasing function of time. Then, we calculated the entropy and internal energy. In order to satisfy second law of thermodynamics, we should consider contribution of dark energy on the total entropy and have shown that the second law of the thermodynamics is valid. Our graphical analysis shows that partition function has Gaussian form in terms of $N$, and for the dark energy proportional to inverse of time, it is a decreasing function of time. The partition function diverges at initial time but it is not a real threat as all physical properties are depending on derivatives of the logarithm of it. It is also clear that the softening parameter reduces the value of the partition function. However, the situation for the Helmholtz free energy is completely opposite which is illustrated by  Fig. \ref{fig2}. There is a minimum of free energy as well as a maximum of entropy for critical $N$ corresponding to the zero-chemical potential. The Helmholtz free energy as well as internal energy are increasing function of time and lead to a constant at the late time. The behavior of the entropy with time is illustrated in the Fig. \ref{fig3} (a). There is a minimum at initial time and it tends to a constant value at the late time. Fig. \ref{fig5} shows the behavior of the specific heat where it tends to positive constant at the late time. Finally, we show that the total entropy is an increasing function of time.

\end{document}